\documentclass[aps,prl,epsfigure,twocolumn,superscriptaddress]{revtex4}
\usepackage{dcolumn}    
\usepackage{color}    
\usepackage{bm} 
\usepackage{graphicx}
\usepackage{amsmath}    
\usepackage{latexsym}
\usepackage{amsfonts}   
\usepackage{amssymb}
\usepackage{array}      
\usepackage{epsfig}
\usepackage{txfonts}
\usepackage[normalem]{ulem}	

\newcommand{\ket}[1]{\left\vert#1\right\rangle}
\newcommand{\bra}[1]{\left\langle#1\right\vert}

\begin{document}
\title{Shortcut to Adiabaticity in the Lipkin-Meshkov-Glick Model}
\author{Steve Campbell}
\affiliation{Centre for Theoretical Atomic, Molecular and Optical Physics, School of Mathematics and Physics, Queen's University, Belfast BT7 1NN, United Kingdom}
\author{Gabriele De Chiara}
\affiliation{Centre for Theoretical Atomic, Molecular and Optical Physics, School of Mathematics and Physics, Queen's University, Belfast BT7 1NN, United Kingdom}
\author{Mauro Paternostro}
\affiliation{Centre for Theoretical Atomic, Molecular and Optical Physics, School of Mathematics and Physics, Queen's University, Belfast BT7 1NN, United Kingdom}
\author{G. Massimo Palma}
\affiliation{NEST, Istituto Nanoscienze-CNR and Dipartimento di Fisica e Chimica, Universit$\grave{a}$  degli Studi di Palermo, via Archirafi 36, I-90123 Palermo, Italy}
\author{Rosario Fazio}
\affiliation{NEST, Scuola Normale Superiore \& Istituto Nanoscienze CNR, I-56126 Pisa, Italy}
\affiliation{Centre for Quantum Technologies, National University of Singapore, 3 Science Drive 2, 117543 Singapore}

\begin{abstract}
We study transitionless quantum driving in an infinite-range many-body system described by the Lipkin-Meshkov-Glick model. Despite the 
correlation length being always infinite the closing of the gap at the critical point makes the driving Hamiltonian of increasing complexity also 
in this case. To this aim we develop a hybrid strategy combining shortcut to adiabaticity and optimal control that allows us to achieve remarkably good performance in suppressing the defect production across the phase transition.
\end{abstract}
\date{\today}
\pacs{} 
\maketitle

The dynamical evolution of a quantum system has often to be tailored so that a given initial state is transformed into 
a suitably chosen target one. In cases such as this, the use of techniques for quantum optimal control can be key in engineering an efficient protocol.
Over the years, formal control methods have been devised, both in the classical and quantum scenario~\cite{krotov}. To date, optimal control has been 
proven beneficial in a multitude of fields, ranging from molecular physics to quantum information processing or high precision measurements~\cite{wamsley}. Only very 
recently, however, this framework has been extended so as to cope with the rich phenomenology and complexity of quantum many-body systems~\cite{montangeroPRL,montangeroPRL2}. In this context, quantum optimal control has been shown to be crucial for the design of schemes for the preparation of many-body quantum states~\cite{montangeroPRL,montangeroPRA, montangeroRAP}, the exploration of the experimentally achievable limits in quantum interferometry~\cite{frank} and the cooling of quantum systems~\cite{chamon}. 

Needless to say, quantum optimal control is not the only way to design the dynamical evolution of a quantum system, and one could consider simpler
(sub-optimal) ways to drive the desired dynamics. For instance, using the adiabatic theorem we are able to constrain a quantum system to remain in an 
eigenstate during any evolution. However, in order for such a technique to be accurate, it should operate on a rather
long timescale. {Unwanted transitions between the state we would like to confine the system into and other ones in its spectrum, which are induced by the unavoidably finite-speed nature of an evolution and ultimately limit the precision of the adiabatic dynamics}, can be 
suppressed by adding suitable corrections to the Hamiltonian guiding the evolution~\cite{berry,rice}. This form of quantum control, named Shortcut To Adiabaticity
(STA), has been considered in a variety of different situations, and recently reviewed in Ref.~\cite{torrontegui}. An experimental implementation using 
cold atomic gases has been reported in Ref.~\cite{bason}.

{Recently, the idea of STA has been extended to quantum many-body systems, a context where it can be potentially very beneficial}. STA has been first employed in the suppression of defects 
produced when crossing a quantum phase transition in {the paradigm model embodied by the} one-dimensional Ising model~\cite{zurek}. {Despite such potential}, a crucial feature that emerges from the use of STA  
in many-body scenarios is the inherent complexity of the (driving) Hamiltonian terms that should be engineered to enable the desired adiabatic process. In fact, it is  the case that the range of the interactions involved in the driving corrections far exceeds that of the model that we aim at controlling: {even $N$-particle models involving two-body interactions (such as the above-mentioned Ising one) require $N$-body driving terms to be run in a fully adiabatic fashion through STA. This obviously makes the implementation of STA in many-body systems quite challenging.} 

{A possible interpretation for the evident complex nature of STA driving terms comes from considering that, close to a critical point of quantum many-body system, the fluctuations of relevant operators of the system, and in turn the correlation length of the model, increase to cover a range far larger than the actual extent of the physical coupling among the particles of the system. Therefore, in order to cancel transitions induced close to the critical point, an $N$-body driving term would be necessary. Notwithstanding the plausible nature of such interpretation, the actual relation between the complexity of the control of a quantum system close to a critical point and the correspondingly diverging correlation length remains yet to be clarified.} In this work we take a 
first, significant step towards the understanding of this important point by considering an infinite range model which undergoes a quantum phase transition. Here the correlation length (which is always infinite) cannot play any role, which would let emerge more neatly the way the peculiarities of a system close to a phase 
transition manifest in the STA. 

{From a pragmatic viewpoint, the complexity of the STA terms often hinder the intuition of the actual mechanism that determines them. {\it Is it possible to design alternative strategies to gather insight into the driving Hamiltonian that realizes the STA}?
By designing a new approach that  combines STA with elements of quantum optimal control inspired from Ref.~\cite{montangeroPRL}, in this work we address the problem of achieving the effectively adiabatic crossing of a critical point in a long-range model.}

To accomplish our goals, we study the Lipkin-Meshkov-Glick (LMG) model~\cite{LMG,vidal1,vidal2,castanos} that was originally posed to study 
shape phase transitions in nuclei and has found fertile application across many fields of physics as a paradigm of an infinitely coordinated 
model. It describes the infinite-range interaction of a set of spin-$1/2$ particles exposed to the effects of an external magnetic field. It 
encompasses the Dicke model, which exhibits a super radiant phase transition, and the Bose-Hubbard model as particular limiting cases 
(we remark achieving STA for the Bose-Hubbard through a different approach was recently studied in Ref.~\cite{bruno} and the adiabatic 
dynamics of the LMG model was examined in Ref.~\cite{fazio}). Simulations of the LMG model have been proposed in systems of circuit quantum electrodynamics~\cite{larson}, single-molecule magnets~\cite{campos} and, very recently, a toroidal Bose-Einstein condensate subjected to the suitable spatial modulation of an external potential~\cite{opatrny}. The latter implementation, in particular, is endowed with sufficient flexibility to implement some of the driving terms that are proposed in this paper.

As we will discuss in details, our hybrid approach provides, in general, driving terms that differ from the prescriptions of STA. Yet, we 
show that we achieve a remarkably good performance when interested in the superadiabatic driving of the LMG model across a 
quantum phase transition and provide a fully constructive method to build corrections for any finite value of $N$, thus going beyond the current state of the art 
in superadiabatic driving of this model~\cite{takahashi}. Remarkably, our 
method does not need full information on the spectrum of the model nor complicated driving potentials, thus lowering the requirements 
for the construction of a STA that approximates accurately the performance of the ideal protocol. Demonstrating the possibility for a 
fully adiabatic crossing of the critical point in such a complex model, which encompasses somehow a `worst-case scenario' in light of 
the infinite range of the interactions being involved, thus demonstrates the full effectiveness of our new approach.

\noindent
{\it Preliminaries.}
The ferromagnetic LMG model  is described by the Hamiltonian
$\mathcal{H}_0(t)=-\frac{1}{N}\left( \sum_{i<j} \sigma_x^i \otimes \sigma_x^j+\gamma \sigma_y^i \otimes \sigma_y^j \right) - 
h(t) \sum_{i} \sigma_z^i$ with $\sigma_{x,y,z}$ the Pauli spin-operators, $h(t)$ the time-dependent magnetic field strength, 
and $\gamma$ the anisotropy parameter. For simplicity, in our simulations we will set $\gamma=0$. However our results are qualitatively unaffected by taking any other value. By considering the collective spin operators $S_\alpha=\sum_i \sigma_{\alpha}^i/2$ 
with $\alpha=\{x,y,z\}$, the model can be written as~\cite{vidal1,vidal2}
\begin{equation}
\label{collspinLMG}
\mathcal{H}_0(t)=-\frac{2}{N}\left( S_x^2 + \gamma S_y^2 \right) - 2h(t) S_z,
\end{equation}
where we have neglected a constant energy shift. The ground state phase diagram consists of two distinct regions and exhibits a 
second order quantum phase transition when $h(t)=1$~\cite{Botet,vidal2}.
In the limit of weak interaction, the LMG model can be solved exactly by mapping it to $N$ bosons in a double well, while in the 
thermodynamic limit, $N\to\infty$, it can be solved through the Holstein-Primakoff (HP) transformation~\cite{vidal1,HP,kwok}. The latter 
approach is also a good approximation for $N\gg1$, although with some limitations~\cite{HP-limits}, and in Ref.~\cite{supp} we illustrate 
this mapping explicitly. 

Following Ref.~\cite{berry} the correction term is calculated from the spectrum of the original Hamiltonian and, 
as will be in our case, there is a choice of phase such that it reads
\begin{equation}
\label{berryeq}
\mathcal{H}_1(t)=i \sum_n \ket{\partial_t \psi_n(t) }\bra{\psi_n(t)},
\end{equation}
where $\ket{\psi_n(t)}$ are the instantaneous eigenstates of $\mathcal{H}_0$. From now on, we omit the time dependence of the parameters and eigenstates. We will assess the performance based on the fidelity $\mathcal{F}=|\bra{\psi_G}{\psi}\big>|^2$ of the evolved state $\ket{\psi}=e^{-i(\mathcal{H}_0+\mathcal{H}_1) t}\ket{\psi_i}$ with the instantaneous ground state $\ket{\psi_G}$ of $\mathcal{H}_0$. We remark that for small $h$, $\ket{\psi_G}$ is not unique as the ground state is degenerate. However, as we are attempting to track the adiabatic dynamics, when we are in this situation we choose one of the two degenerate states as our ground state. This sets our analysis apart from that of Ref.~\cite{bruno}, where the model considered was closely related to the antiferromagnetic LMG and no such degeneracy occurs.

\begin{figure}[t]
{\bf (a)}\hskip4cm{\bf (b)}\\
\includegraphics[width=0.48\columnwidth]{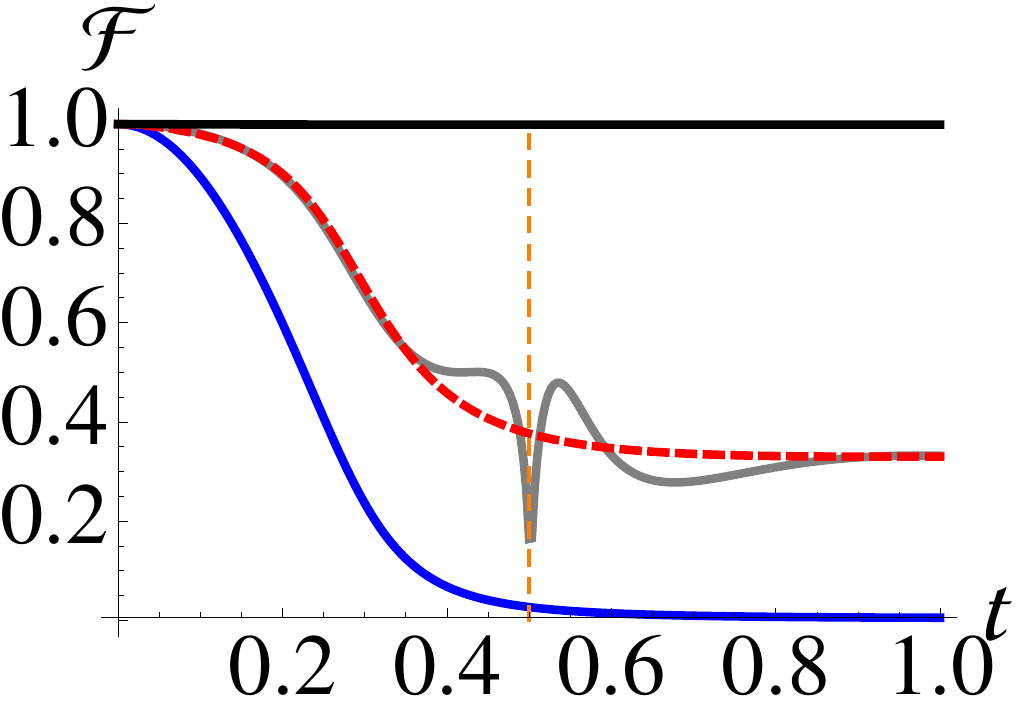}~~\includegraphics[width=0.48\columnwidth]{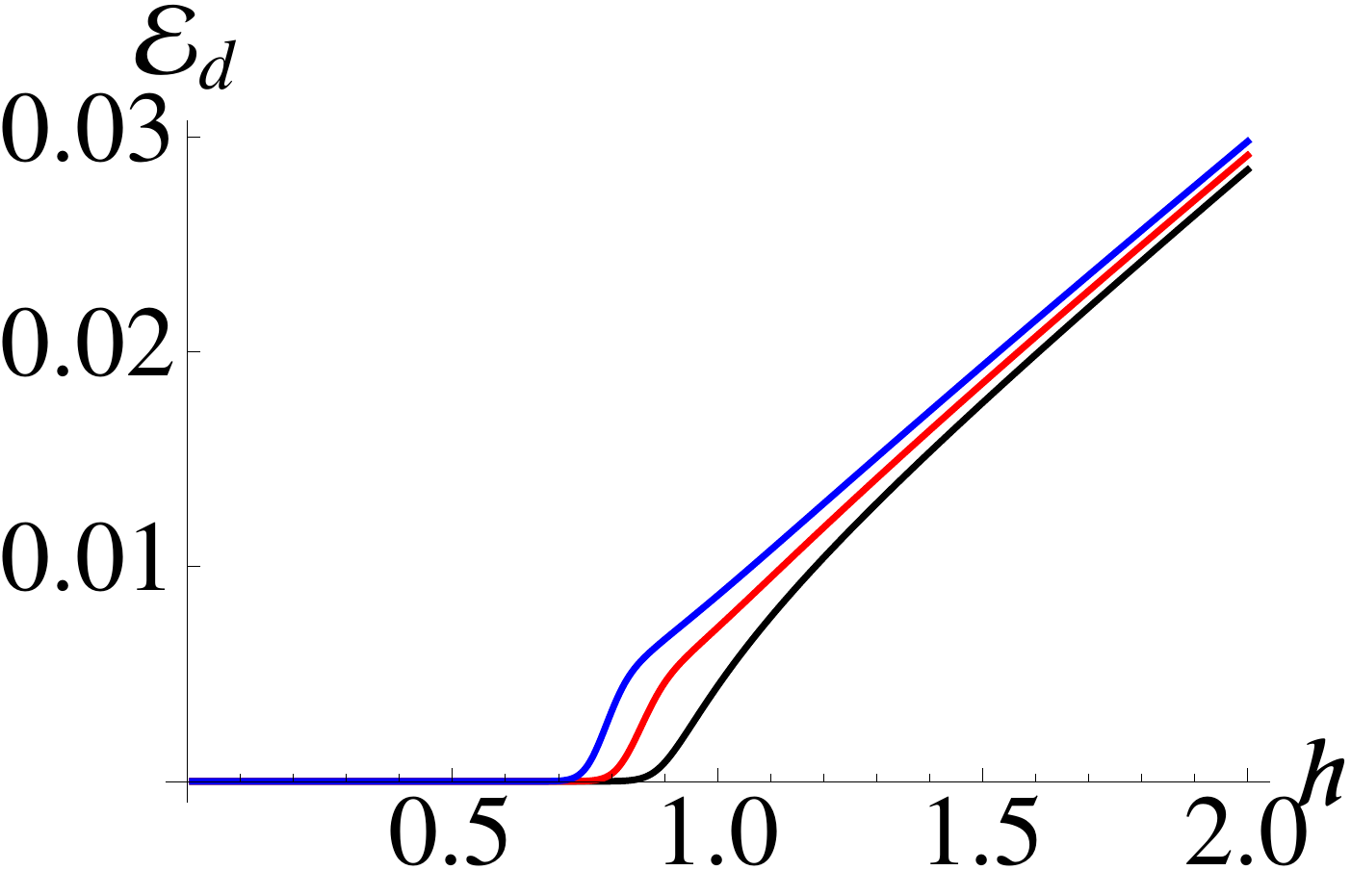}
\caption{(Color online) {\bf (a)} Fidelity of the evolved states with the instantaneous ground state for the full numerically calculated correction term [topmost, black line], the truncated version [dashed red line], the harmonic oscillator correction [gray line], and the bare Hamiltonian for a linear ramp [blue line], for $h(t)=0.75+0.5t$. The dashed vertical line at $t=0.5$ corresponds to the critical point $h=1$. {\bf (b)} Energy difference, $\mathcal{E}_d$, between the ground and first excited [right-most black curve], second and third excited [middle red curve], and fourth and fifth excited [left most blue curve] energy levels of $\mathcal{H}_0$ against field strength, $h$. For large $h$ we see the energy levels are distinct, however as $h$ is decreased they become pairwise degenerate. In both panels $N=100$ and $\gamma=0$.}
\label{fig1}
\end{figure}

\noindent
{\it Approach 1: Direct calculation.} For small $N$ we can readily calculate the correction term. Working in the basis of maximum angular momentum (which is a constant of motion) and using the eigenstates of $S_z$ labelled as $\ket{0},\dots\ket{N}$, we can diagonalize Eq.~\eqref{collspinLMG} and find the corresponding eigenstates. For $N=2$ these are $\ket{\psi_1} = \sin\theta\ket{2} + \cos\theta\ket{0}$, $\ket{\psi_2}=\ket{1}$, and $\ket{\psi_3}=\cos\theta\ket{2} - \sin\theta\ket{0}$, with $\theta$ a function of $h$ and $\gamma$. Clearly we have 2 distinct subspaces that are never mixed, meaning that our correction term is effectively that of a single two-level system~\cite{mugaPRL}. In terms of the collective spin operators we find $\mathcal{H}_1=\dot{\theta}~(S_x S_y + S_y S_x )$. Going on, for $N=3$ we can analytically determine the shortcut, once again observing that the Hamiltonian establishes two distinct subspaces, each spanned by two eigenstates. The correction term is then effectively that of two independent two level systems. In terms of collective spin operators, its expression can be found following the method described in Ref.~\cite{supp} as $\mathcal{H}_1=[2({\dot{\theta}_1+\dot{\theta}_2})B_1+({\dot{\theta}_1-\dot{\theta}_2}) B_2]/\sqrt3$ with $B_1=S_x S_y + S_y S_x$ and $B_2= S_x S_y S_z + S_z S_y S_x$. When written in terms of the single-spin operators, $B_{1,2}$  describe two and three-body couplings among the particles of the system. Physically, it is straightforward to check that $B_{1,2}$ conserve the parity of the spin-system state, that is $[B_{1,2},\Pi_{e,o}]=0$ with $\Pi_e$ ($\Pi_o$) the projector onto the subspace with an even (odd) number of excitations. Despite such symmetry, the analytic assessment of Eq.~\eqref{berryeq} becomes intractable for $N>3$.
However, by addressing numerically the cases of $N=4,..,10$ it is possible to extrapolate the following general form (which we conjecture to be valid for any size of the system)
\begin{equation}
\label{fullSC}
\mathcal{H}_1=i\left(
\begin{array}{cccccccc}
0 & 0 & -x_{1,1} & 0 & -x_{2,1} & 0 & -x_{3,1} & \dots \\
0 & 0 & 0 & -x_{1,2} & 0 & -x_{2,2} & 0 &  \\
x_{1,1} & 0 & 0 & 0 &  -x_{1,3} & 0 &  -x_{2,3} &  \\
\vdots &  &   &  &  & \ddots & &  \\
\end{array}
\right).
\end{equation}
Here, the coefficients $x_{i,j}$'s stand for the coefficients of the decomposition of the $j^{\rm{th}}$ eigenstate of ${\cal H}_0$ over the chosen basis. On one hand, this implies that the construction of ${\cal H}_1$ would require the knowledge of the whole spectrum of the model under scrutiny. On the other hand, the correction term for an $N$-spin problem would require coupling operators involving up to $N$ spins. The engineering of such driving term thus appears to be daunting.
Upon inspection, the set $\{x_{1,j}\}$ is found to be orders of magnitude larger than the other elements entering Eq.~\eqref{fullSC}. This leads us to conjecture that $x_{1,j}$'s are dominant in the correction. By forcefully suppressing all other elements, the fidelity with the instantaneous ground state using this approach, when we vary the magnetic field as $h(t)=0.75+0.5t$ (here $t$ is a dimensionless time) for $N=100$, is shown by the dashed red curve in Fig.~\ref{fig1} {\bf (a)}. This is a change of the magnetic field strength (fast with respect to the natural evolution time of the system, which would be of the order of $N$), that takes the system across its critical point at $h=1$. Similar conclusions are reached for other forms of the ramping magnetic field strengths, in~\cite{supp} we examine some further examples. 

The performance of these approximate shortcuts can be well understood by considering the energy spectrum of Eq.~\eqref{collspinLMG}. In Fig.~\ref{fig1} {\bf (b)} we examine the lowest six energy levels. When $h$ is large all energy levels are uniquely defined. However, below a threshold value (that approaches 1 for $N\to\infty$), the spectrum becomes pairwise degenerate. Starting from the ground state for $h<1$, as is the case in Fig.~\ref{fig1} {\bf (a)}, we are in a region where the energy levels are degenerate. This entails that, without the full correction term, transitions are likely to occur quickly due to the vanishing gap between levels. Starting from the opposite phase, i.e. $h>1$, we find that all approaches (including the bare Hamiltonian) perform significantly better until we approach the degeneracy point~\cite{supp}.

\begin{figure}[b]
\includegraphics[width=0.6\columnwidth]{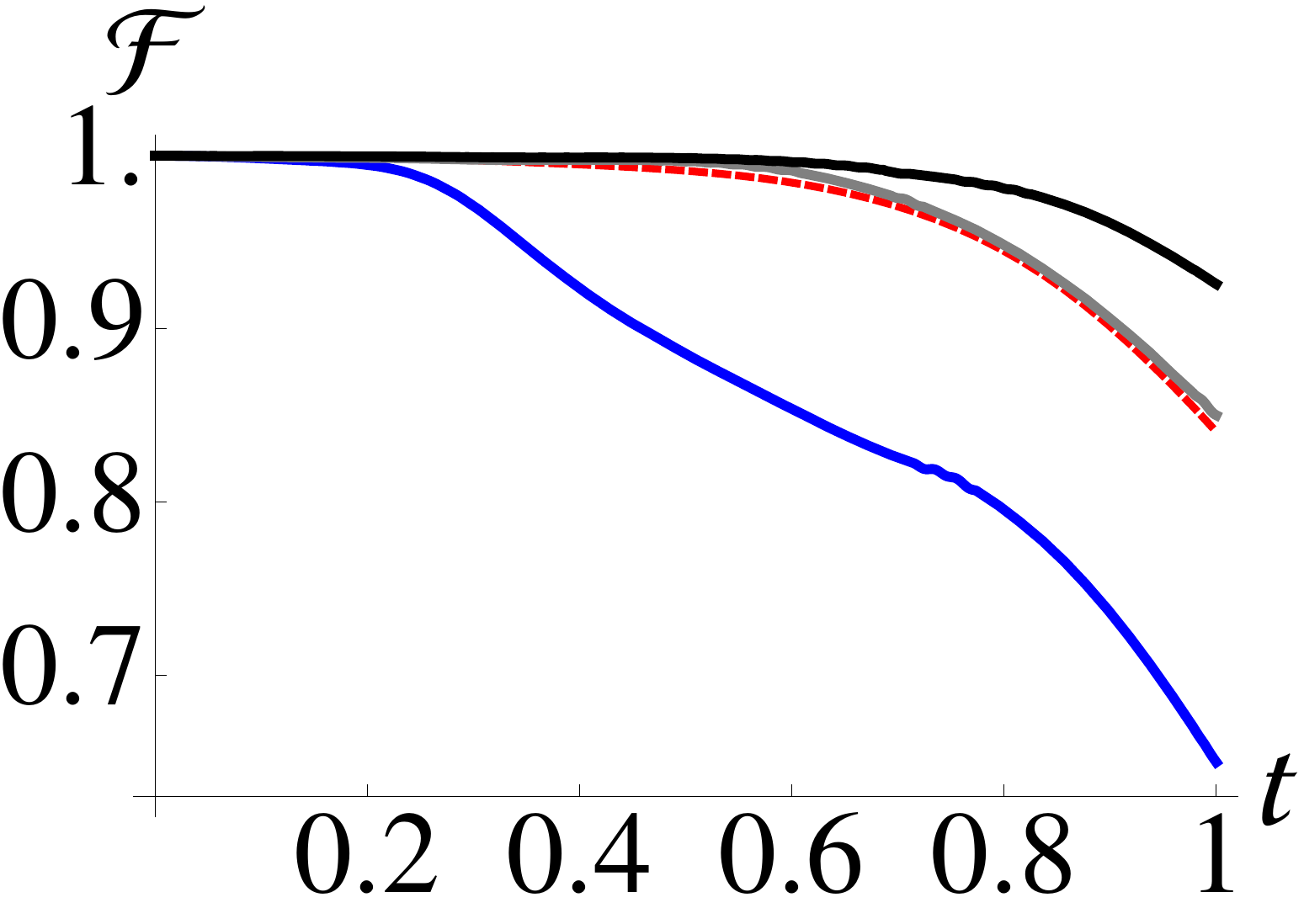}
\caption{(Color online) Fidelity of the evolved state using the `band' structured correction term Eq.~\eqref{Hprime} including the first one, two, three and four bands going from bottom to top. We consider the same linear ramp as for Fig.~\ref{fig1} and take $\gamma=0$ and $N=80$.}
\label{fig2}
\end{figure}

\noindent
{\it Approach 2: Ansatz optimization.}
Based on our analytical and numerical results we can deduce that while for full STA we require a complete knowledge of the spectrum, even using suboptimal approaches can give significant improvements over the bare Hamiltonian. Based on the analysis in the previous Section, we now know that regardless of system size the correction term populates diagonal `bands', the leading of which closely resembles $S_xS_y+S_yS_x$. As the previous Section clearly showed, a driving term populating only these elements can lead to a dramatic increase in performance over a simple linear ramp. However, while the truncated Hamiltonian would appear simpler, determining the corresponding elements in principle still requires the knowledge of all the eigenstates of $\mathcal{H}_0$. We thus turn our attention to the anticipated hybrid approach involving STA and optimal control-type techniques. In essence, we aim to achieve the best possible performance without requiring the complete knowledge of the spectrum of the original Hamiltonian. We achieve this by assuming that our driving Hamiltonian is populated in diagonal bands, similarly to Eq.~(\ref{fullSC}), as 
\begin{equation}
\label{Hprime}
\mathcal{H}_1'=i\left(
\begin{array}{cccccccc}
0 & 0 & -x_{1} & 0 & -x_{2} & 0 & -x_{3} & \dots \\
0 & 0 & 0 & -x_{1} & 0 & -x_{2} & 0 & \\
x_{1} & 0 & 0 & 0 &  -x_{1} & 0 &  -x_{2} &  \\
\vdots &  &   &  &  & \ddots &  &  \\
\end{array}
\right).
\end{equation}
Noticeably, this conjectured form of the driving term still belongs to the class of parity-preserving Hamiltonians. We can then solve the system's Schr\"odinger equation
$
i\bra{k}\partial_t\ket{\psi(t,\{x_i\})} = \bra{k}  \mathcal{H}_0+\mathcal{H}_1' \ket{\psi(t,\{x_i\})} 
$
with $k$ running over the eigenstates of $S_z$ and the initial condition $\ket{\psi(0,\{x_i\})}=\ket{\psi_G}$. The corresponding solutions can then be optimized to find the values of $\{x_i\}$ that maximize $\mathcal{F}$ at all instants of time. In Fig.~\ref{fig2} we show the performance for $N=80$ with $h(t)=0.75+0.5t$. The lowest curve is when only the first band is considered, and each successively higher curve corresponds to an additional band being included. Quite remarkably with a single band we find a significant increase in the performance over all previous approaches and including just four is sufficient to achieve a fidelity of $\mathcal{F}>0.92$. 

\begin{figure}[t]
{\bf (a)}\hskip4cm{\bf (b)}\\
\includegraphics[width=0.48\columnwidth]{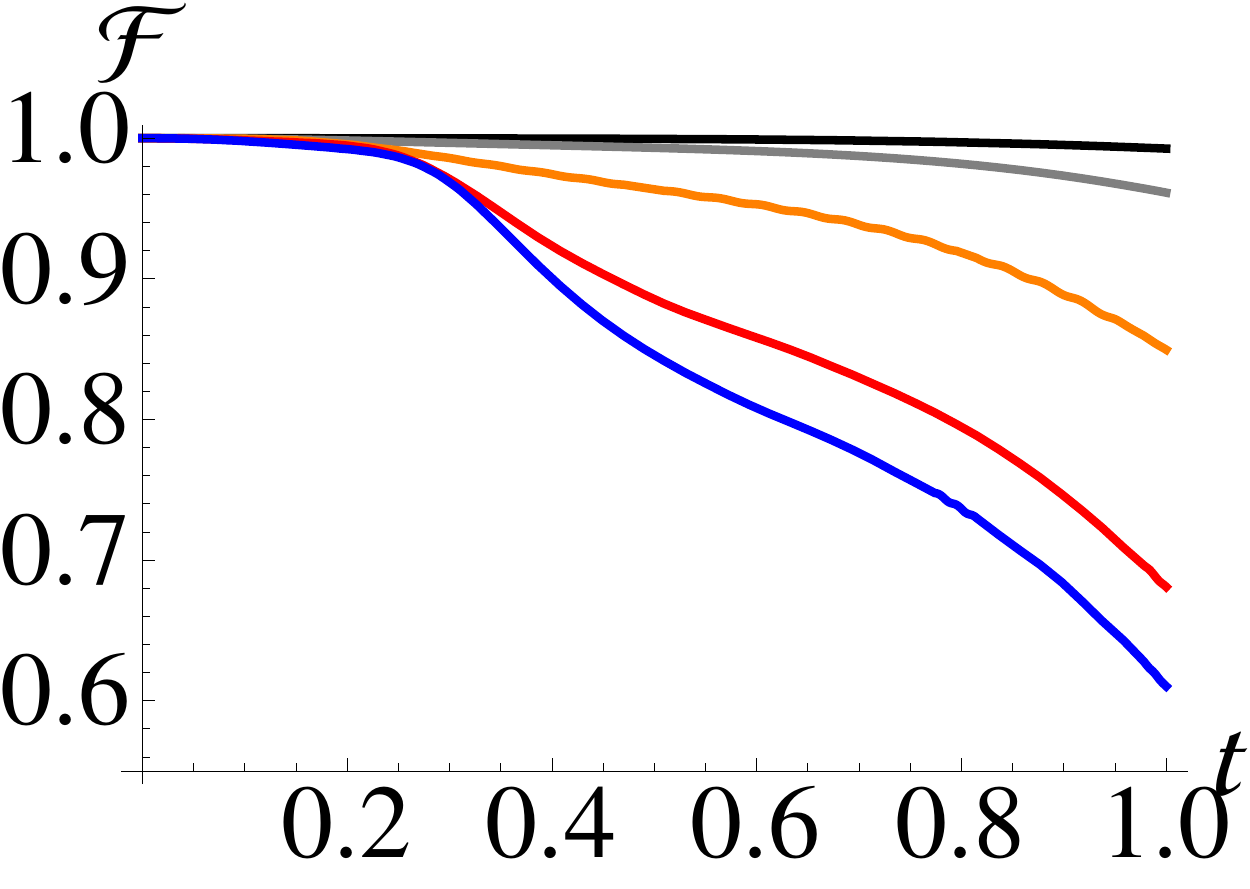}~~~\includegraphics[width=0.48\columnwidth]{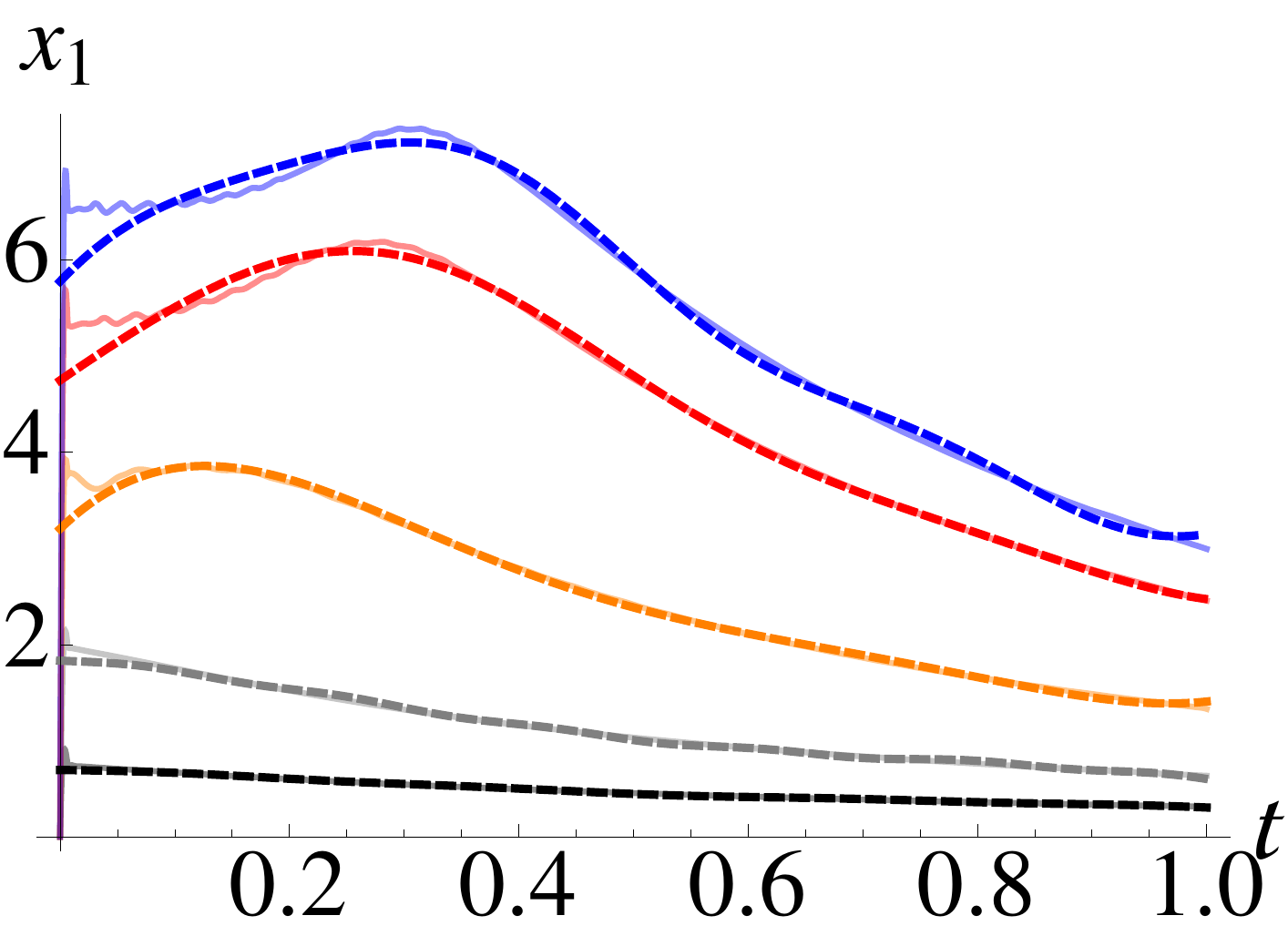}
\caption{(Color online)  {\bf (a)} Fidelity of evolved state with instantaneous ground state for $N=10$, 20, 40, 80, and 100, top to bottom respectively, when only the first band in Eq.~\eqref{Hprime} is considered for the linear ramp $h(t)=0.75+0.5t$. {\bf (b)} The light solid curves are the numerically optimized values of $x_1$ for $N$ increasing from bottom to top. The dashed curves are the results of the harmonic fits $x^f_1$. In both panels $\gamma=0$.}
\label{fig3}
\end{figure}

Focusing on the first band, in Fig.~\ref{fig3} we see how this approach scales as we increase $N$. For small systems, $N\sim10$, this technique maintains a fidelity of $\mathcal{F}>0.99$. Increasing $N$ we still maintain fidelities much larger than the bare Hamiltonian, however bigger systems will require more bands to be included as we go through the transition point to maintain a high performance due to the significant increase in the Hilbert space. In Fig.~\ref{fig3} {\bf (b)} the solid curves correspond to the numerically optimized values for $x_1$, while the dashed lines are the harmonic series fit $x_1^f=\sum_{m=1}^3 a_m\sin(\omega_m t + \phi_m)$. Including only three harmonics is already sufficient to closely approximate the optimized $x_1$, and the quantitative difference in $\mathcal{F}$ by using this functional form is negligible~\cite{supp}. This suggests that this hybrid approach is quite robust to small fluctuations in the pulse shape, as is clear on examination of Fig.~\ref{fig3} {\bf (b)}~\cite{montangeroPRL3}.

While the decomposition of Eq.~\eqref{fullSC} in terms of physical operators appears extremely difficult, we can find a decomposition for each band that allows us to build up the full correction term. Indeed, from the previous analysis we know the first band is dominant, as indicated by the biggest increase in fidelity over the bare Hamiltonian. For any finite $N$ we can construct it as $\mathcal{H}_1^{x_{1}}= \sum_{i=1}^{N-1}\sum_{j=0}^{N-2} x_{1,i}~\beta_{i,j}~B_j$, where
\begin{equation}
\label{band}
B_j = \begin{cases}
S_z^{\frac{j}{2}} (S_x S_y +S_y S_x)  S_z^{\frac{j}{2}} ~&j~\text{even}, \\
  S_z^{\frac{j-1}{2}}  S_x S_y  S_z^{\frac{j+1}{2}}  + S_z^{\frac{j+1}{2}}  S_y S_x  S_z^{\frac{j-1}{2}} ~&j~\text{odd}.
 \end{cases}
\end{equation}
with the coefficients $\beta_{i,j}$ being directly calculated (cf. Ref.~\cite{supp}). Including more bands is somewhat more involved, although we have been able to devise a constructive method to build them, as illustrated in Ref.~\cite{supp}. This allows us to clearly see the physical resources necessary to implement the full shortcut in Eq.~\eqref{fullSC} and the band structured version of Eq.~\eqref{Hprime}. In particular, we note from Eqs.~\eqref{band} that all terms are constructed based on $(S_xS_y+S_yS_x)$. We see that our method allows us to increase the performance to a desired level while keeping the necessary resources as simple (comparative to the complexity of the original Hamiltonian) as possible. 

\noindent
{\it Comparison with Holstein-Primakoff mapping.} We now compare the results achieved through our hybrid approach to what is gathered in the large $N$ limit by exploiting the HP transformation. Regardless of the phase the system is in, we find that we can map it to a harmonic oscillator as
\begin{equation}
\mathcal{H}_\text{ho}= \begin{cases}
                                    2\sqrt{(h-1)(h-\gamma)}  \left( b^\dagger b +\frac{1}{2}  \right) - h ( N+1 ),~h>1,\\
                                   2\sqrt{(1-h^2)(1-\gamma)}  \left( b^\dagger b +\frac{1}{2}  \right) - \frac{1+h^2}{2} N - \frac{1-\gamma}{2},~h\in(0,1).
                                   \end{cases}
\end{equation}
In~\cite{mugaJPB} the corresponding correction term for the simple harmonic oscillator with time dependent frequency, $\omega(t)$, was calculated, $\mathcal{H}_\text{ho}=i\hbar [\partial_t\ln{\omega(t)}] ( b^2 - b^{\dagger^2})/4$, with $b~(b^\dagger)$ the annihilation (creation) operators for the harmonic oscillator mode, and leads us to the following driving Hamiltonian
$\tilde{\mathcal{H}}_{1}=f(h,\gamma)( S_xS_y + S_yS_x )$
with $f(h,\gamma)$ given in Ref.~\cite{supp}. However, this correction term is not freed from issues: it is not defined at the transition point $h=1$ and is exact only when $N\to\infty$. Therefore, for any finite value of $N$ we can expect some unwanted transitions to occur. In Fig.~\ref{fig1} {\bf (a)} the gray curve corresponds to the fidelity of the evolved state using this shortcut with the instantaneous ground state when we linearly vary the field as $h(t)=0.75+0.5t$ for $N=100$. The discontinuity at $t=0.5$ is due to $\tilde{\cal H}_1$ not being defined at this point, therefore we simply assume the correction term is `switched off' as it transitions into the new phase. Although we do not achieve perfect STA, we see a remarkable increase in the values taken by the fidelity compared to the bare Hamiltonian evolution (lowest blue curve). The sub-optimal performance of this shortcut can be traced back to the fact that for {\it any} finite $N$ the HP transformation gives only an approximation. 

\noindent
{\it Conclusions.} 
We have examined the conditions to achieve full transitionless quantum driving  for the LMG model. When dealing with finite $N>3$ we found we must explicitly calculate the correction term numerically. By examining the structure of the resulting Hamiltonian we were able to develop a hybrid approach to achieve remarkably good performance by employing optimization to an ansatz constructed by examining the numerical and analytical forms calculated previously. Even for large systems this allows for a significant simplification on the requirements for achieving near perfect STA by not necessitating the complete knowledge of the spectrum. The complexity in the control of a system close to a phase transition thus goes beyond the fact that the range of the correlation is diverging. In our opinion, our work on the STA in the LMG clearly identifies this point showing that it is the critical slowing down (and the related closing of the gap) the source of complexity in implementing STA in critical systems. In the limit of $N\to\infty$, the model can be mapped to a harmonic oscillator through the Holstein-Primakoff transformation and we found the corresponding correction term takes a simple form proportional to $B_1$ regardless of the phase, however this correction term is not defined at the critical point. Additionally, due to the limitations on the validity of the HP transformation for large (but finite $N$) the harmonic oscillator approximation fails to achieve a high performance. This approach holds the potential to fruitful applications in situations of high physical relevance based on the physics of quantum many-body systems. As an interesting example, it could be employed to devise low-entropy protocols for the extraction of work from quantum spin systems driven out of equilibrium without (or with significantly quenched) concomitant friction, which is a key transformation in micro- and nanoscale machines~\cite{engines}, or to achieve highly entangled multiparticle ground states with effectively adiabatic protocols operating at finite time and limited entropic byproducts.

The main challenge in the context of the proposal put forward here is the physical implementation of the driving term that would guarantee the STA-like dynamics of the system. While this is a characteristic that is common to STA-based protocols in quantum many-body systems~\cite{delcampo}, we believe that the approach discussed here will be key in achieving an experimental proof-of-principle. A seemingly potential candidate system could be the one put forward in Ref.~\cite{opatrny}, where Hamiltonian terms of the form of ${B}_{1,2}$ for instance can be engineered.

{\it Acknowledgements.}
The authors are grateful to A. del Campo, L. Fusco, M.-\'A. Garc\'ia-March, C. Jarzynski, B. Juli\'a-D\'iaz, S. Montangero, and D. Poletti for useful discussions. This work is funded by the EU Collaborative Project TherMiQ (Grant Agreement 618074), EU Project IP-SIQS, the UK EPSRC (through grants EP/G004579/1, EP/L005026/1, and EP/K029371/1), the John Templeton Foundation (grant ID 43467), and the Italian PRIN-MIUR. We acknowledge partial support from the COST Action MP1209.

\renewcommand{\theequation}{S-\arabic{equation}}
\setcounter{equation}{0}  

\section{Supplementary Material}
\section{Mapping to the harmonic oscillator}
In this section we outline solving the LMG model, as considered previously in~\cite{vidal1,kwok}. Starting from Eq.~(1) in the main text (including the constant energy shift for consistency with the literature)
\begin{equation}
\label{collspinLMG}
\mathcal{H}_0=-\frac{2}{N}\left( S_x^2 + \gamma S_y^2 \right) - 2h S_z + \frac{\lambda(1+\gamma)}{2}.
\end{equation}
For $N\to\infty$, the model can be solved through the Holstein-Primakoff (HP) transformation that allows us to map the spin model to an equivalent harmonic oscillator. Care must be taken however, depending on the phase, the HP transformation must be taken along the direction that the classical angular momentum, 
\begin{equation}
\text{{\bf S}}=\frac{N}{2}(\sin\varphi\cos\phi,\sin\varphi\sin\phi,\cos\varphi),
\end{equation}
points. For $h>1$ we find this is always along the $z$-axis. Neglecting terms higher than $O(N)$ the HP transformation in this limit is
\begin{equation}
\label{HP1}
S_+=\sqrt{N} a, ~~~~~S_-=\sqrt{N} a^\dagger,~~~~~S_z=\frac{N}{2}-a^\dagger a,
\end{equation}
with
\begin{equation}
\label{HP2}
S_x=\frac{1}{2}(S_+ + S_-)~~~~~\text{and}~~~~S_y=\frac{1}{2i}(S_+ - S_-).
\end{equation}
This results in the mapped Hamiltonian in terms of bosonic creation and annihilation operators
\begin{equation}
\mathcal{H}_b=-\frac{1-\gamma}{2} \left( a^2 + a^{\dagger~2} \right) + (2h-1-\gamma) a^\dagger a - hN,
\end{equation}
which can then be written in diagonal form by performing the following Bogoliubov transformation
\begin{eqnarray}
a&=\sinh\left(\frac{\alpha}{2}\right) b^\dagger + \cosh\left(\frac{\alpha}{2}\right) b,\\
a^\dagger&=\sinh\left(\frac{\alpha}{2}\right) b + \cosh\left(\frac{\alpha}{2}\right) b^\dagger,
\end{eqnarray}
and taking 
\begin{equation}
\tanh \alpha = \frac{1-\gamma}{2h-1-\gamma},
\end{equation}
we finally obtain the harmonic oscillator equivalent for our Eq.~\eqref{collspinLMG}
\begin{equation}
\label{mappedLMGlargeh}
\mathcal{H}_\text{ho}= 2\sqrt{(h-1)(h-\gamma)}  \left( b^\dagger b +\frac{1}{2}  \right) - h ( N+1 ) + \frac{1+\gamma}{2}.
\end{equation}

For $0<h<1$ this classical vector moves as $h$ is varied. Therefore before performing the HP transformation the Hamiltonian must be rotated to be inline with the direction of the classical angular momentum, or equivalently we must take the HP transformation along the direction this vector points for a given value of $h$. We shall take the latter approach.

For clarity, let us look at the slightly simpler case of $\gamma=0$. In this case the classical vector moves between pointing along the $x$-axis ($h=0$) and pointing along the $z$-axis ($h=1$) according to $\varphi=\arccos h$. Therefore, we take the HP transformation along this new direction
\begin{equation*}
\begin{aligned}
\mathcal{H}^\varphi &= -\left(\frac{2}{N} \right) \left(S_x^{\varphi}\right)^2 -2h S_z^\varphi +\frac{1+\gamma}{2},\\
S_x^\varphi &=  S_x \cos\varphi  - S_z \sin\varphi ,\\
S_z^\varphi  &=  S_z \cos\varphi  + S_x\sin\varphi .
\end{aligned}
\end{equation*}
We now use the same operators as in Eqs.~\eqref{HP1} and \eqref{HP2} and therefore we have no need to perform any inverse rotations after the mapping is complete. Doing this results in a different bosonic representation,
\begin{equation}
\mathcal{H}_b=-\frac{1}{2}h^2 \left(a^2+a^{\dagger~2}\right)+\left(2-h^2\right) a^{\dagger}a- \left( \frac{h^2N}{2}+\frac{h}{2} +\frac{N}{2}  \right)+\frac{1}{2}.
\end{equation}
Taking
\begin{equation}
\tanh \alpha = \frac{h^2}{2-h^2},
\end{equation}
in the Bogoliubov operators, we obtain the harmonic oscillator equivalent for $\gamma=0$ 
\begin{equation}
\mathcal{H}_\text{ho}=2\sqrt{(1-h^2)}  \left( b^\dagger b +\frac{1}{2}  \right) - \frac{1+h^2}{2} N - \frac{1}{2}.
\end{equation}
When considering arbitrary $\gamma$, the calculation is slightly more involved. However the final form achieved is
\begin{equation}
\label{mappedLMGsmallh}
\mathcal{H}_\text{ho}=2\sqrt{(1-h^2)(1-\gamma)}  \left( b^\dagger b +\frac{1}{2}  \right) - \frac{1+h^2}{2} N - \frac{1-\gamma}{2}.
\end{equation}

\section{Driving Hamiltonian for $N\to\infty$}
Here we give the explicit form of the correction term when $N\to\infty$. From~\cite{mugaJPB}, for the `standard' harmonic oscillator
\begin{equation}
\mathcal{O}_0=\hbar \omega \left( b^\dagger b + \frac{1}{2}  \right)
\end{equation}
the driving Hamiltonian is 
\begin{equation}
\mathcal{O}_{1}=i\hbar \frac{\Dot{\omega}}{4 \omega} \left( b^2 - b^{\dagger^2}  \right).
\end{equation}
For our purposes, working in units of $\hbar=1$, we take the effective frequency term appearing in Eqs.~\eqref{mappedLMGlargeh} and \eqref{mappedLMGsmallh} as $\omega$, i.e.
\begin{equation}
\omega=\begin{cases}
                   2\sqrt{(h-1)(h-\gamma)}, ~~~&h>1,\\
                   2\sqrt{(1-h^2)(1-\gamma)},~~~&0<h<1,
                   \end{cases}
\end{equation}
remembering that $h$ is time dependent. Given that $\left( a^2 - a^{\dagger^2}  \right) = \left( b^2 - b^{\dagger^2}  \right)$, and returning to the collective spin operators, we can then determine the correction term to be
\begin{equation}
\mathcal{H}_{1}=\begin{cases}
                               \frac{2h-1-\gamma}{4N(h-1)(h-\gamma)}\left( S_xS_y + S_yS_x  \right),~~~h>1,\\
                               \frac{2h(\gamma-1)}{4N(1-h^2)(1-\gamma)}\left( S_xS_y + S_yS_x  \right),~~~0<h<1.
                               \end{cases}
\end{equation}

\section{Finite $N$ driving Hamiltonian in terms of collective spin operators}
In this section we examine the structure of the driving Hamiltonian for finite $N$ in terms of the physical operators required to construct it. The correction for $N=3$ is given by
\begin{equation}
\mathcal{H}_1=i\left(
\begin{array}{cccc}
0 & 0 & -\dot{\theta}_1 & 0 \\
 0 & 0 & 0 & -\dot{\theta}_2 \\
\dot{\theta}_1 & 0 & 0 &0 \\
0 & \dot{\theta}_2  & 0 &0 
\end{array}
\right).
\end{equation}
We can express this in terms of the collective spin operators using Eqs.~(6) and $\tilde{\cal H}_1$ from the main text. Taking $N=3$ and noting that $x_{1,i}=\dot{\theta}_i$, Eq~(6) becomes
\begin{equation}
\label{betaeq}
\mathcal{H}_1=\dot{\theta}_1 \left( \beta_{1,0} B_0 +   \beta_{1,1} B_1 \right) + \dot{\theta}_2 \left( \beta_{2,0} B_0 +   \beta_{2,1} B_1 \right),
\end{equation}
with
\begin{eqnarray*}
B_0&=&S_xS_y+S_yS_x,\\
B_1&=&S_xS_yS_z+S_zS_yS_x.
\end{eqnarray*}
We can then determine $\beta_{i,j}$ by equating $\tilde{\cal H}_1$ with
\begin{equation}
\left(
\begin{array}{cccc}
0 & 0 & -i & 0 \\
 0 & 0 & 0 & -i \\
i & 0 & 0 &0 \\
0 &i  & 0 &0 
\end{array}
\right),
\end{equation}
finding
\begin{equation}
\beta_{1,0}=\beta_{2,0}=\frac{1}{2\sqrt{3}},\qquad \beta_{1,1}=-\beta_{2,1}=\frac{1}{\sqrt{3}}.
\end{equation}
We thus have 
\begin{equation}
\begin{split}
\mathcal{H}_1=&\frac{\dot{\theta}_1}{2\sqrt{3}} \left[ ( S_x S_y + S_y S_x ) + 2 ( S_x S_y S_z + S_z S_y S_x ) \right]  \\
                            +&   \frac{\dot{\theta}_2}{2\sqrt{3}} \left[ ( S_x S_y + S_y S_x ) - 2 ( S_x S_y S_z + S_z S_y S_x ) \right] .
\end{split}
\end{equation}
This approach can be applied {\it verbatim} for any $N$ to determine the first band in terms of the collective spin operators. Indeed, for $N=4$, from Eq.~(4) we know that the driving Hamiltonian takes the form
\begin{equation}
\mathcal{H}_1=i\left(
\begin{array}{ccccc}
0 & 0 & -x_{1,1} & 0 & -x_{2,1} \\
 0 & 0 & 0 & -x_{1,2} & 0 \\
x_{1,1} & 0 & 0 & 0 & -x_{1,3} \\
0 & x_{1,2}  & 0 &0 & 0 \\
x_{2,1} & 0 & x_{1,3}  &0 & 0 
\end{array}
\right),
\end{equation}
which can be expressed in terms of the collective spin operators as
\begin{equation}
\begin{split}
\mathcal{H}_1&=  \frac{x_{1,1}+x_{1,3}}{2\sqrt{6}} \left(S_xS_y+S_yS_x\right) + \frac{x_{1,1}-x_{1,3}}{2\sqrt{6}} \left(S_xS_yS_z+S_zS_yS_x\right)\\ 
                            &+ \left( \frac{x_{1,1}+x_{1,3}}{2\sqrt{6}} - \frac{x_{1,2}}{3} \right) \left(  S_zS_xS_yS_z+S_zS_yS_xS_z  \right)\\
                            &+ \frac{i x_{2,1}}{24} \left( S_-^4 - S_+^4  \right).
\end{split} 
\end{equation}
We see that the recipe for constructing the first band is the same as described for $N=3$. The second band is found be noticing that for band $b$, the terms populated are proportional to $i\left( S_-^{2b}-S_+^{2b} \right)$. By employing similar techniques that lead to Eqs. (6) and $\tilde{\cal H}_1$ of the main text, we can determine their explicit form as well.

\begin{figure}[t]
{\bf (a) \hskip0.4\columnwidth (b)}\\
\includegraphics[width=0.45\columnwidth]{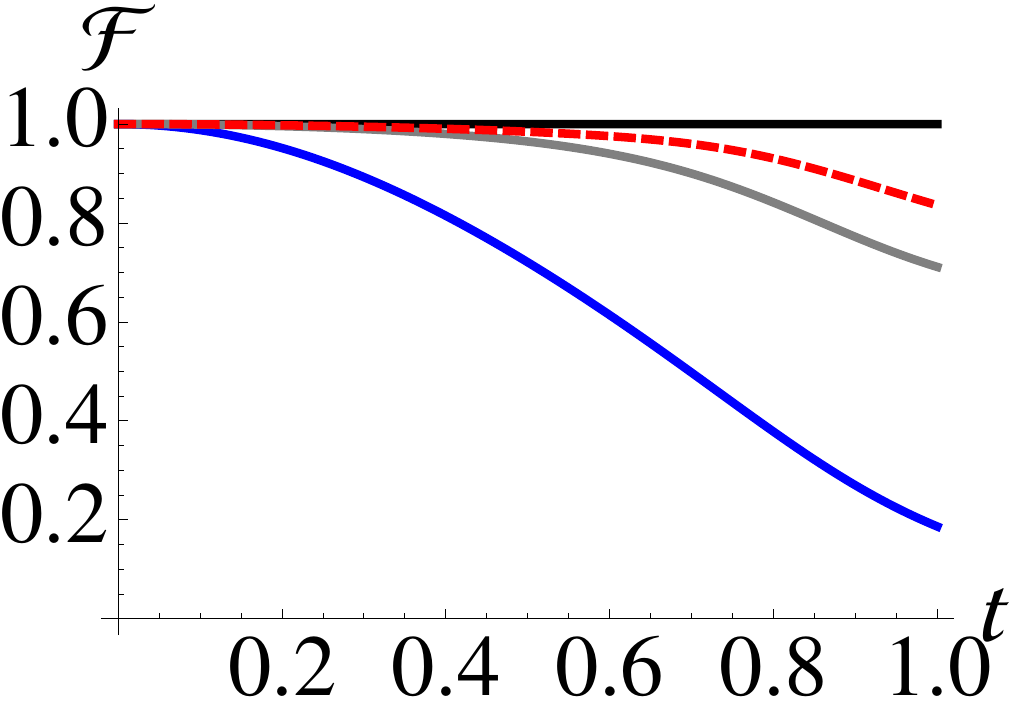}~~\includegraphics[width=0.45\columnwidth]{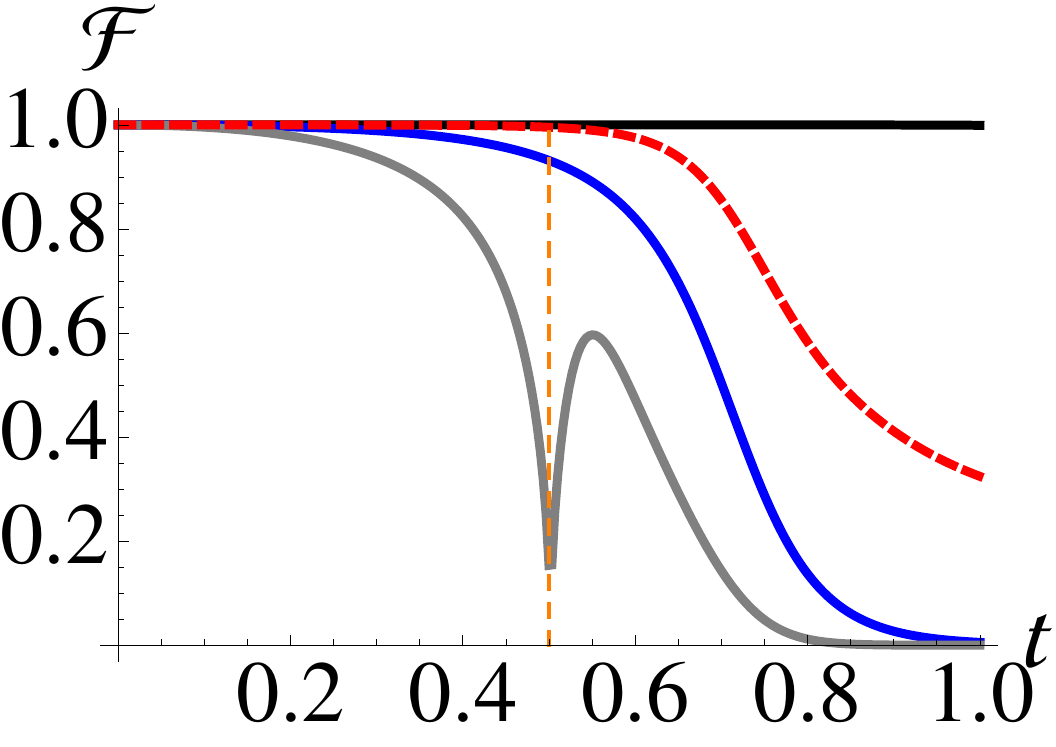}\\
{\bf (c)\hskip0.4\columnwidth (d)}\\
\includegraphics[width=0.45\columnwidth]{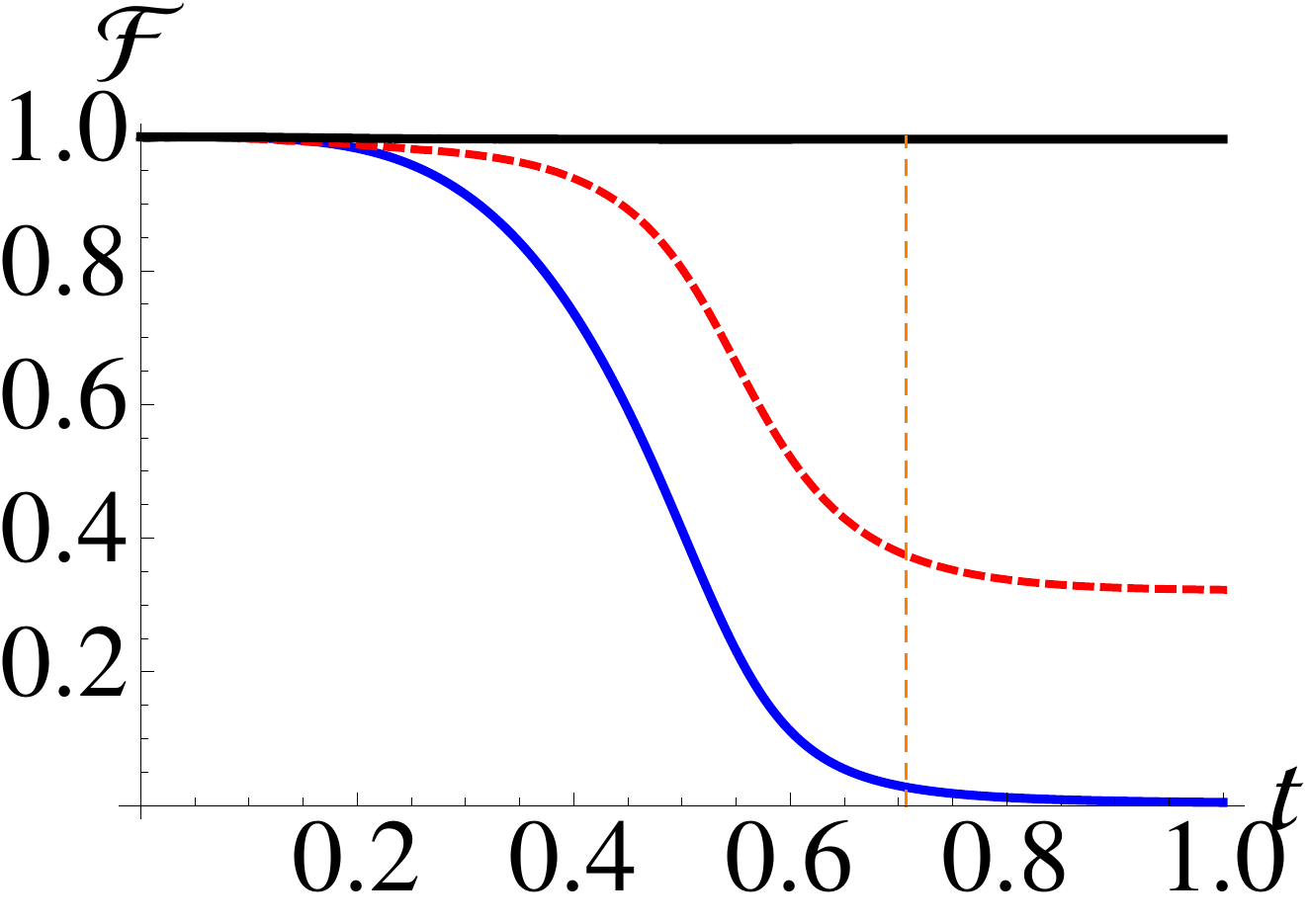}~~\includegraphics[width=0.45\columnwidth]{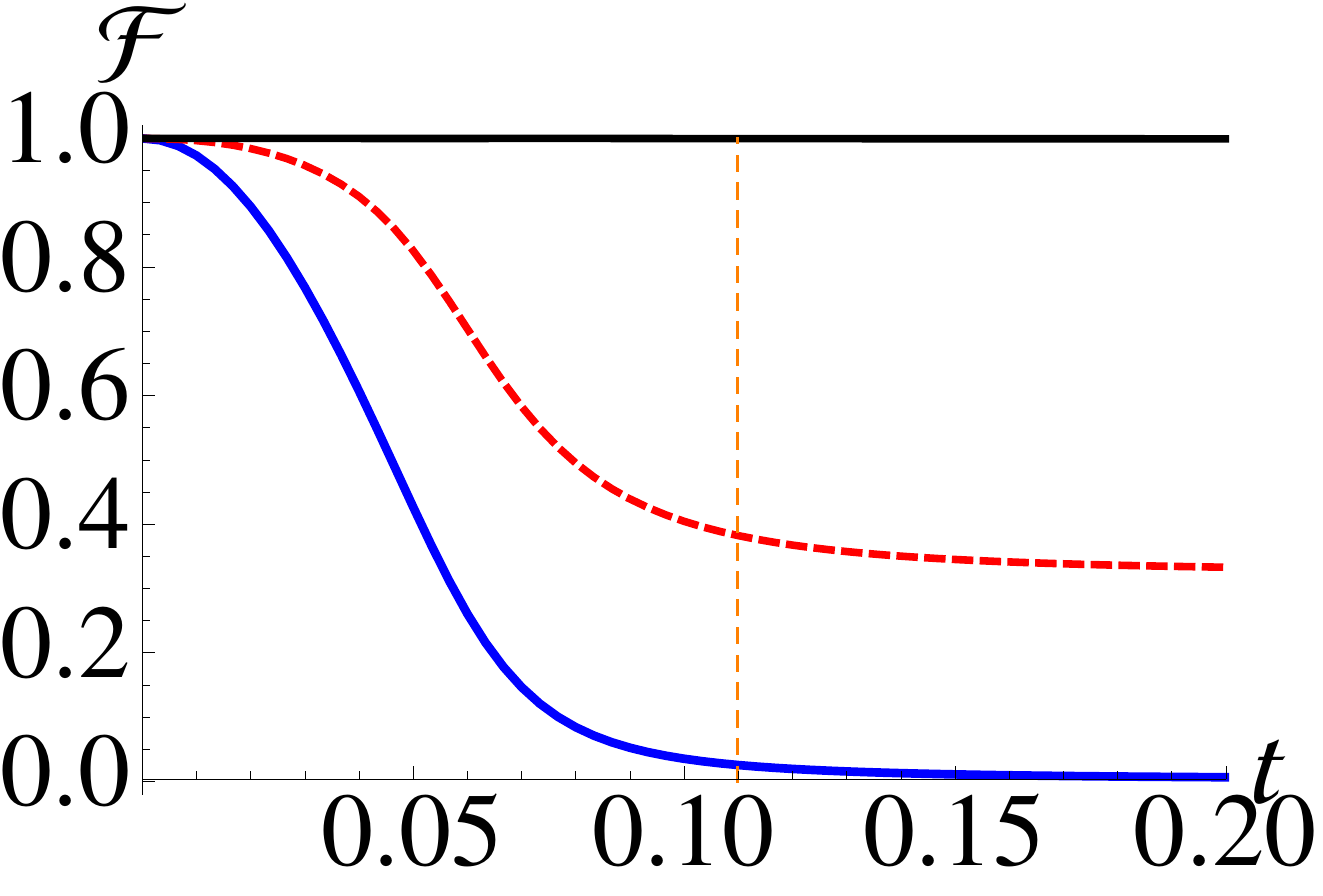}
\caption{ In all panels we examine the fidelity with the instantaneous ground state for the full shortcut [topmost, black line], truncated correction [dashed, red curve], harmonic approximation [gray curve], and bare hamiltonian [blue curve] when $N=100$ for various different ramps. {\bf (a)} Linearly varying $h(t)=0.55+0.3t$ therefore restricting to one phase of the model. {\bf (b)} Linearly varying $h(t)=1.25-0.5t$, i.e. ramping $h$ exactly contrary to that in the main text. {\bf (c)} Quadratically varying $h(t)=0.75+0.5 t^2$ {\bf (d)} Varying the field as $h(t)=0.75+0.5 \tanh(5t)$. The dashed vertical lines in panels {\bf (b)}-{\bf (d)} correspond to the values of $t$ when the ramp reaches the critical point $h=1$.}
\label{fig}
\end{figure}

\section{Performance of correction terms: Additional examples}
To illustrate more completely the behavior of the various approaches we include additional examples. In Fig.~\ref{fig} {\bf (a)} we examine the performance of the full, truncated, harmonic approximation, and bare hamiltonian for a different linear ramp, $h(t)=0.55+0.3t$. In this case we do not go through the transition, and remain in the pairwise degenerate phase of the model. As clearly demonstrated, while all approaches surpass the bare hamiltonian we see, the truncated form performs significantly better than the harmonic approximation. In panel {\bf (b)} we reverse the direction of the ramp used in the main text: $h(t)=1.25-0.5t$. Now a number of features are apparent, notice all correction terms perform significantly better until $t$ approaches 0.5 (i.e. $h\to1$). This increased performance is due to the large gap between the ground and excited states, meaning unwanted transitions are less likely occur. Additionally, we now see the harmonic approximation is consistently poorer than the bare Hamiltonian. This is again, in part, due to the increased separation between energy levels initially: the greater difference in energy implies that doing nothing, i.e. bare hamiltonian, is in fact quite close to optimal, while the harmonic approximation correction causes more significant changes to the evolution. Finally, in panel {\bf (c)}  we quadratically vary the field strength $h(t)=0.75+0.5 t^2$ and in {\bf (d)}  $h(t)=0.75+0.5 \tanh(5t)$. In both these examples we see that the qualitative behavior is largely unaffected by changing the way in which the variation in the field is performed, with the performance achieved though our proposal providing substantial improvements over the bare dynamics. 

\section{Scaling of the fidelity with the number of harmonics}
Here we examine how accurate the harmonic series expansion is with the numerically optimized pulses. As stated in the main text, the harmonic fit is given by
$x_1^f=\sum_{m=1}^c a_m\sin(\omega_m t + \phi_m).$
Table~\ref{table1} shows the maximum discrepancy between the fidelity when the optimized value of $x_1$ is taken with the fidelity when the harmonic series fit value is used, $\left(\text{i.e.}~\mathcal{F}_{x_1}-\mathcal{F}_{x_1^f}\right)$. Already for two harmonics the difference is of the order of $10^{-3}$, and for larger systems three harmonics approximate the ideal behavior excellently. Furthermore, the small differences observed in $\mathcal{F}$ clearly show that even with small fluctuations in the shape of the applied pulse, we still achieve a consistent performance. 
\begin{table}[h!]
\centering
\begin{tabular}{|c||c|c|c|}
\hline 
~~$N$~~&~~$c=1$~~&~~$c=2$~~&~~$c=3$~~\\ [0.5ex]
\hline 
~~10~~&~~0.005~~&~~0.002~~&~~0.002~~\\
~~40~~&~~0.024~~&~~0.008~~&~~0.003~~\\
~~70~~&~~0.112~~&~~0.002~~&~~0.0003~~\\
\hline
\end{tabular}
\caption{Difference between the fidelity calculated using the numerically optimized values of $x_1$ and the values of $\mathcal{F}$ obtained using a decomposition in $c$ harmonics according to the series fit.}
\label{table1}
\end{table}

\end{document}